\documentclass[preprint,titlepage,aps,prd,
tightenlines,amsmath,byrevtex,nofootinbib]{revtex4}

\usepackage{graphicx}

\begin{document}

\title{
Probing Non Standard Neutrino Physics at T2KK and Neutrino Factory
}

\author{Shoichi Uchinami}
\email{uchinami@phys.metro-u.ac.jp}

\affiliation{Department of Physics, Tokyo Metropolitan University, 
1-1 Minami-Osawa, Hachioji, Tokyo 192-0397, Japan}

\begin{abstract}
We discuss the possibility of constraining or discovering the non-standard neutrino physics 
beyond the standard model with future long baseline neutrino oscillation 
experiments. 
Among very many possible experimental set up we discuss neutrino factory and 
T2KK, in which two detector setting may be useful to achieve the goal. In particular, neutrino factory 
with two detectors at baselines of 3000 km and 7000 km have a great sensitivity to non-standard 
interaction (NSI) and solve the $\theta_{13}$-NSI confusion problem.

\end{abstract}

\maketitle

\section{Introduction}
In order to determine the neutrino oscillation parameters there is a lot of future 
experimental projects which would have ability of precision measurement. It means that 
we may have sensitivity to not only the standard oscillation parameters but also 
some new physics effect on neutrino sector. In this talk, we discuss about probing 
the non-standard neutrino physics with future long baseline experiment based on 
the works\cite{NSI-nufact}\cite{NSP-T2KK}.

Due to limitations of space, we would like to concentrate mainly on the work of non-standard 
neutrino interaction with neutrino factory.

\section{Non-Standard neutrino Interaction with neutrino factory}
Since early time, it is suggested the possible existence of non-standard 
interaction(NSI) of neutrinos with matter\cite{wolfenstein}\cite{valle}\cite{guzzo}. 
In this section we discuss about probing such a NSI with future long baseline 
experiments.

Introducing the lagrangian for NSI:
\begin{eqnarray}
        {\cal L}_{\mbox{eff}}^{\mbox{NSI}} = 
        -2 \sqrt{2} \varepsilon_{\alpha\beta}^{fP} G_F
        (\overline{\nu}_\alpha \gamma_\mu P_L \nu_\beta)\,
        (\overline{f} \gamma^\mu P f),
\label{NSI-Lag}
\end{eqnarray}
where $G_{F}$ is the Fermi constant, $f$ for the species of fermions in the earth 
e, u, d, and $P^{L}_{R} \equiv \frac{1}{2}(1 \mp \gamma_{5})$. 
In this notation, $\varepsilon_{\alpha \beta}^{fP}$ is the effective coupling normalized 
by week interaction.

Concentrating on the effect of NSI in neutrino propagation, the evolution equation is 
written as,

\begin{eqnarray} 
i {d\over dt} 
        \left( \begin{array}{c} 
        \nu_e \\ \nu_\mu \\ \nu_\tau 
        \end{array}  \right)
 = 
        \left[ U \left( \begin{array}{ccc}
                   0   & 0          & 0   \\
                   0   & \frac{\Delta m^2_{21}}{2 E}  & 0  \\
                   0   & 0           &  \frac{\Delta m^2_{31}}{2 E}  
                   \end{array} \right) U^{\dagger} +  
        a \left( \begin{array}{ccc}
                1 + \varepsilon_{ee}   & \varepsilon_{e\mu} & \varepsilon_{e\tau} \\
                \varepsilon_{e \mu }^* & \varepsilon_{\mu\mu} & \varepsilon_{\mu\tau} \\
                \varepsilon_{e \tau}^* & \varepsilon_{\mu \tau }^* & \varepsilon_{\tau\tau} 
        \end{array} \right) \right] ~
        \left( \begin{array}{c} 
                   \nu_e \\ \nu_\mu \\ \nu_\tau 
        \end{array}  \right)
\label{NSI-evolution}
\end{eqnarray}
where U is the Maki-Nakagawa-Sakata flavor mixing matrix\cite{MNS}, $a \equiv \sqrt{2} G_{F} n_{e}$, $n_{e}$ is the electron number density in the 
earth, and $\varepsilon_{\alpha \beta}$ describes the total effect of NSI coming from charged fermion e, u, d. 
Though off-diagonal $\varepsilon$ parameters may have complex phase, we assume in this talk that all $\varepsilon_{\alpha \beta}$'s are real.

In order to improve the sensitivity to NSI, we examine the power of golden channel 
at neutrino factory\cite{Nufact}. Neutrino factory has a great sensitivity to $\sin^2 2 \theta_{13}$. 
Considering that 
$\varepsilon_{\alpha \beta}$ produces non-standard flavor changing effect, it is natural to expect 
that we have a high sensitivity also to NSI parameters. In fact, the contribution to 
oscillation probability $P(\nu_{e} \to \nu_{\mu})$ of 
$\varepsilon_{e \mu}$ and $\varepsilon_{e \tau}$ are same order to the one of 
$\sin \theta_{13}$. 
Though this feature helps us to probe the NSI, 
but at the same time possible existence of NSI may destroy the precision 
measurement of $\theta_{13}$ and CP-phase $\delta$, the so called confusion problem
\cite{confusion1}\cite{confusion2}.

\begin{figure}[htb]
\begin{center}
        \includegraphics[scale=0.6]{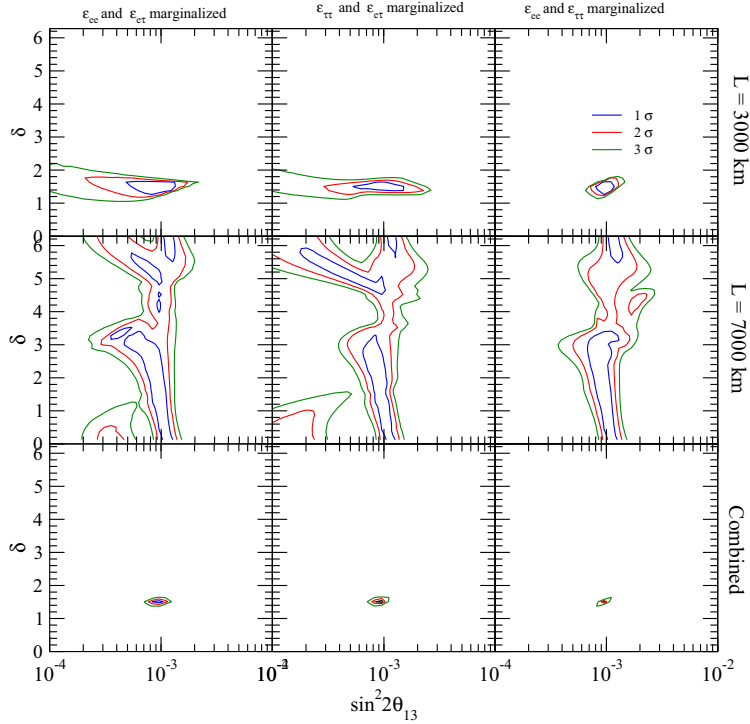}
        \caption{Allowed regions of $\sin^22 \theta_{13}$ and CP-phase $\delta$ with use of the $\nu_{e} \to \nu_{\mu}$ and $\bar{\nu}_{e} \to \bar{\nu}_{\mu}$ 
channel for various combination of NSI elements $\varepsilon_{\alpha \beta}$. The upper(middle) panels are for a detector 
at 3000(7000)km, and the bottom panels for the combined one. 
 The input values of $\varepsilon_{\alpha \beta}$ are zero, 
$\sin^2 2\theta_{13}=0.001$, and $\delta=\pi/2$.
 The figure complements those in \cite{NSI-nufact} with $\delta=\pi/4$ and $3\pi/2$.
}
\label{fig_th13del_const}
\end{center}
\end{figure}

See the upper panels of Figure \ref{fig_th13del_const}. This is the allowed region of 
standard oscillation parameters $\sin^2 2 \theta_{13}$ and $\delta$ on neutrino factory 
with a detector at 3000 km obtained by marginalizing certain combinations of two $\varepsilon_{\alpha \beta}$'s.
As noted above, the sensitivity to the standard oscillation parameters become significantly worse. 
In particular we cannot determine the nonzero value 
of $\theta_{13}$ even at the input value of $\sin^2 2 \theta_{13}=10^{-3}$ 
for the case with $\epsilon_{e \tau}$.

In order to solve this confusion problem, we consider the 2nd detector at 7000 km.
This is nearby so called magic baseline\cite{huber-winter}\cite{smirnov}, where $L=7200$ km 
for constant earth matter density $\rho=4.5$g/cc. 
The magic baseline is the special distance because $P(\nu_{e} \to \nu_{\mu})$ is independent 
of the solar oscillation parameters and $\delta$. At the magic baseline, oscillation 
probability with $\varepsilon_{e \tau}$ is written as\cite{NSI-nufact}:
\begin{eqnarray}
        P(\nu_{e} \to \nu_{\mu}:magic BL)
        =
        4 s_{23}^2 \biggl |s_{13} e^{-i \delta} \frac{\Delta_{31}}{a} + 
          c_{23} \varepsilon_{e \tau} \biggr |^2 
        \left( \frac{a}{\Delta_{31} - a} \right)^2
        \sin^2 \left( \frac{\Delta_{31} - a}{2} L \right)
        \label{P_MBL}
\end{eqnarray}
where $s_{ij} \equiv \sin \theta_{ij}$, $c_{ij} \equiv \cos \theta_{ij}$, and $\Delta_{31} \equiv \frac{\Delta m_{31}}{2E}$. With the information of complex phase $\delta$ at 3000 km, we can determine a certain combination of $\theta_{13}$ and $\varepsilon_{e \tau}$ at the magic baseline. 
This advantage helps us solve the confusion problem. Bottom panels in Figure \ref{fig_th13del_const} 
are the allowed region of $\sin^2 2\theta_{13}$ and $\delta$ with the combined information of 
two detectors. The sensitivity becomes dramatically higher.

\begin{figure}[htb]
\begin{center}
        \includegraphics[scale=0.7]{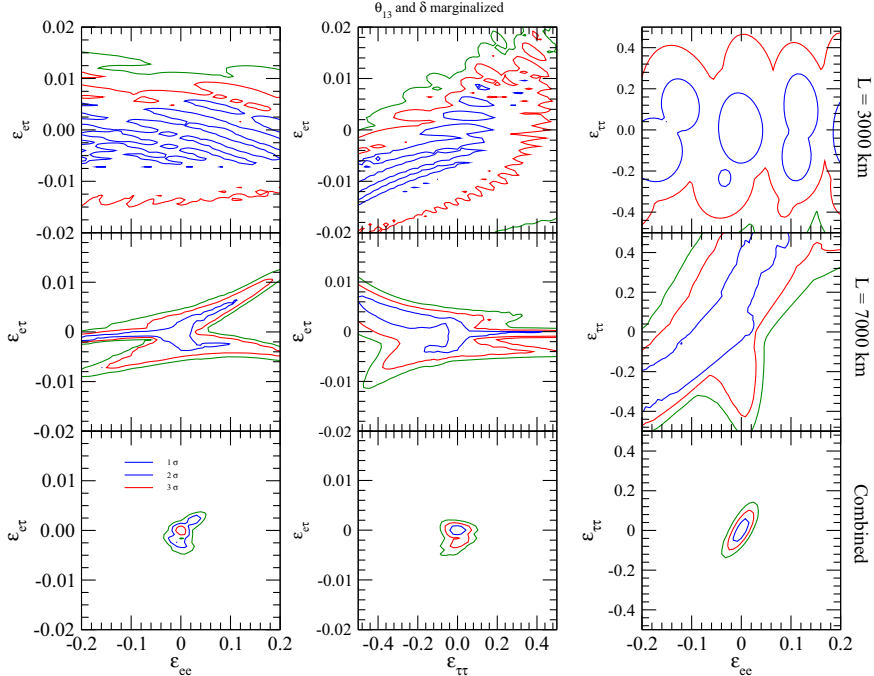}
        \caption{Allowed regions of various combination of two NSI parameters. 
                 The input values are same as 
Figure \protect\ref{fig_th13del_const}.
}
\label{fig_eps_const}
\end{center}
\end{figure}

The synergy effect is also powerful for probing the NSI. Figure \ref{fig_eps_const} 
presented the allowed region of the NSI parameters. Though it is very hard to determine the $\epsilon_{\alpha \beta}$ 
with a individual detector, we can determine the NSI parameters 
very accurately if we combine the two.

In this work\cite{NSI-nufact}, the sensitivity reach for $\varepsilon_{\alpha \beta}$ we have obtained is:
\begin{eqnarray}
        |\varepsilon_{e \tau}| \simeq O(10^{-3})~,~|\varepsilon_{e \mu}| \simeq O(10^{-4}).
\end{eqnarray}
In more recent work with information of $\nu_{\mu}$ disappearance channel\cite{kopp}, 
the sensitivity reach to NSI parameters in $\nu_{\mu}-\nu_{\tau}$ sector is reported as
\begin{eqnarray}
        |\varepsilon_{\mu \tau}| \simeq O(10^{-4}), 
        |\varepsilon_{\mu \mu}|,|\varepsilon_{\tau \tau}| \simeq O(10^{-2}).
\end{eqnarray}
They confirmed that the two detector setting at $L=4000$ km and $L=7500$ km, 
which is similar to ours, is close to optimal even if NSI in matter are considerd. 
Notice that it is nothing but the standard IDS-NF setup\cite{ISS}\cite{IDS-NF}.
\

Finally we would like to point out that even though the true value of $\theta_{13}$ vanishes 
or extremely small, 
in which case we cannot measure the CP-phase $\delta$ neither, 
we would have a reasonable 
sensitivity to non-standard neutrino interaction. 
It is because 
the contribution of $\varepsilon_{e \mu}$ and $\varepsilon_{e \tau}$ to 
$P(\nu_{e} \to \nu_{\mu})$ remains as 
$\varepsilon_{e \mu,\tau} \times \frac{\Delta m_{21}^2}{2Ea}$ and 
$|\varepsilon_{e \mu,\tau}|^2$ terms in the perturbative formula.

\section{Non-Standard neutrino Physics with T2KK}
In the relatively near future experiment, we have the sensitivity to NSI on the $\mu \tau$ 
sector $|\varepsilon_{\mu \tau}|$ and $|\varepsilon_{\tau \tau}|$.
We briefly mention hare the potential of T2KK\cite{T2KK} for searching the NSI using $\nu_{\mu}$ 
disappearance channel\cite{NSP-T2KK}.

For small $\varepsilon_{\alpha \beta}$, $\nu_{\mu}$ disappearance channel can be 
written by two flavor approximation. In such system, the effects of NSI to oscillation 
probability are large with $\varepsilon_{\mu \tau}$, $\varepsilon_{\mu \mu}$ and, 
$\varepsilon_{\tau \tau}$. So concentrating these $\varepsilon$'s, the sensitivity 
at 2 $\sigma$ CL. may be reached by T2KK setup:
\begin{eqnarray}
        |\varepsilon_{\mu \tau}| \lesssim 0.03(0.03),~~~~
        |\varepsilon_{\tau \tau}| \lesssim 0.3~(1.2).
\label{T2KK-NSI_const}
\end{eqnarray}
where $\sin^2 \theta = 0.45(0.5)$ and $\varepsilon_{\mu \mu}$ is assumed zero because 
only the difference can be measured.

\section{Conclusion}
We discussed the power of the two detector setting at different baseline distance 
for probing non-standard neutrino physics. 
Neutrino factory, especially with two detectors at 3000 and 7000 km, is a powerful 
setting for solving the confusion problem that make the precision measurement of standard 
oscillation parameters untenable if we consider the possible existence of NSI. 
It is also useful to search the NSI even for extremely small value of NSI parameters.

\begin{acknowledgements}
I am very grateful to my collaborators, Hisakazu Minakata, Hiroshi Nunokawa, Renata 
Zukanovich Funchal, and Nei Cipriano Ribeiro for valuable collaboration. 
Osamu Yasuda gave me a fruitful comment. 
This work was supported in part by Grant-in-Aid for JSPS Fellows No. 209677.
\end{acknowledgements}

\end{document}